\documentclass[5p]{elsarticle}

\usepackage{hyperref}
\usepackage{listings}
\usepackage{graphicx}
\usepackage{amsmath}
\usepackage{upgreek}
\usepackage{sansmath}

\graphicspath{{Figures/}}

\begin{document}

\title{Optimized digital filtering techniques for radiation detection with HPGe detectors}
\author[rvt]{Marco Salathe\corref{cor1}}
\ead{marco.salathe@mpi-hd.mpg.de}
\author[rvt]{Thomas Kihm}
\ead{mizzi@mpi-hd.mpg.de}

\cortext[cor1]{Corresponding author}
\address[rvt]{Max-Planck-Institut f{\"u}r Kernphysik, Saupfercheckweg 1, 69117 Heidelberg, Germany}

\begin{abstract}
This paper describes state-of-the-art digital filtering techniques that are part of GEANA, an automatic data analysis software used for the GERDA experiment. The discussed filters include a novel, nonlinear correction method for ballistic deficits, which is combined with one of three shaping filters: a pseudo-Gaussian, a modified trapezoidal, or a modified cusp filter. The performance of the filters is demonstrated with a 762\,g Broad Energy Germanium (BEGe) detector, produced by Canberra, that measures $\gamma$-ray lines from radioactive sources in an energy range between 59.5 and 2614.5\,keV. At 1332.5\,keV, together with the ballistic deficit correction method, all filters produce a comparable energy resolution of $\sim1.61$\,keV FWHM. This value is superior to those measured by the manufacturer and those found in publications with detectors of a similar design and mass. At 59.5\,keV, the modified cusp filter without a ballistic deficit correction produced the best result, with an energy resolution of 0.46\,keV. It is observed that the loss in resolution by using a constant shaping time over the entire energy range is small when using the ballistic deficit correction method.
\end{abstract}
\begin{keyword}
Digital filtering techniques \sep ballistic deficit correction \sep high purity germanium detectors (HPGe) \sep radiation detection \sep $\gamma$-ray spectroscopy \sep GERDA 
\end{keyword}

\maketitle

\section{Introduction}
\label{sec:Intro}
\label{sec:dsp}
The use of High Purity Germanium (HPGe) detectors is reliant on the application of effective signal filtering methods to properly estimate the amount of energy deposited by radiation. Notably, the evaluation of the data quality for experiments with multiple detectors requires fast and accurate digital signal processing routines. GEANA ("GErmanium ANAlysis") is a novel digital signal processing software created to address this need in the scope of the GERmanium Detector Array (GERDA) experiment \cite{GerdaExp2013}. It is capable of monitoring the steady flow of data collected by the experiment over several years. 

The energy deposited by radiation in the germanium detectors is reconstructed by processing the digitized signals with digital filters. Digital filters compensate for effects, such as the decay of the signal induced by the RC-feedback of the charge sensitive preamplifier, the finite charge collection time of the signal and frequency components dominated by noise. This publication aims to describe the filters used to counter these effects within GEANA. 

Shaping filters are used to improve the signal-to-noise ratio. The pseudo-Gaussian filter, the most common digital shaping filter, brings the input pulse close to a Gaussian distribution. This shaping filter is fast in execution, simple to implement and its performance is sufficient for many applications. However, for particular situations other filters perform more optimally. The current signal of a typical HPGe detector, unfortunately, does not have the shape of a delta function. Nevertheless, if it is assumed that a delta function is an input signal to a circuit that only adds current and voltage noise, the infinite cusp filter \cite[p.~245]{fluoresence}, has the best performance. However, the infinite nature of the cusp filter conflicts with the finite length of digital signals. 

The time profile of the current signal (pulse shape), that is induced into the read out electronics varies with the number and location of energy deposition sites in a HPGe detector. Neither the cusp nor the pseudo-Gaussian filter are well suited to deal with variations in the pulse shape. These filters underestimate the amount of energy deposited in the detector for pulses with long rise times relative to pulses with a short rise time. This effect is called ballistic deficit and can be partially compensated for by choosing a sufficiently large shaping time or the trapezoidal filter \cite{Radeka1972}. 

In the past, two additional approaches were proposed to correct ballistic deficits with analog modules. The first approach calculated a semi-empirical compensation which depended on the input signal rise time and then was added to the final output signal \cite{Goulding}. The second approach used a gated integrator after prefiltering the signal to integrate over the charge collection time \cite{Moszynski}. 

\section{Methods}
\label{sec:meth}
\begin{figure}
\centering \includegraphics{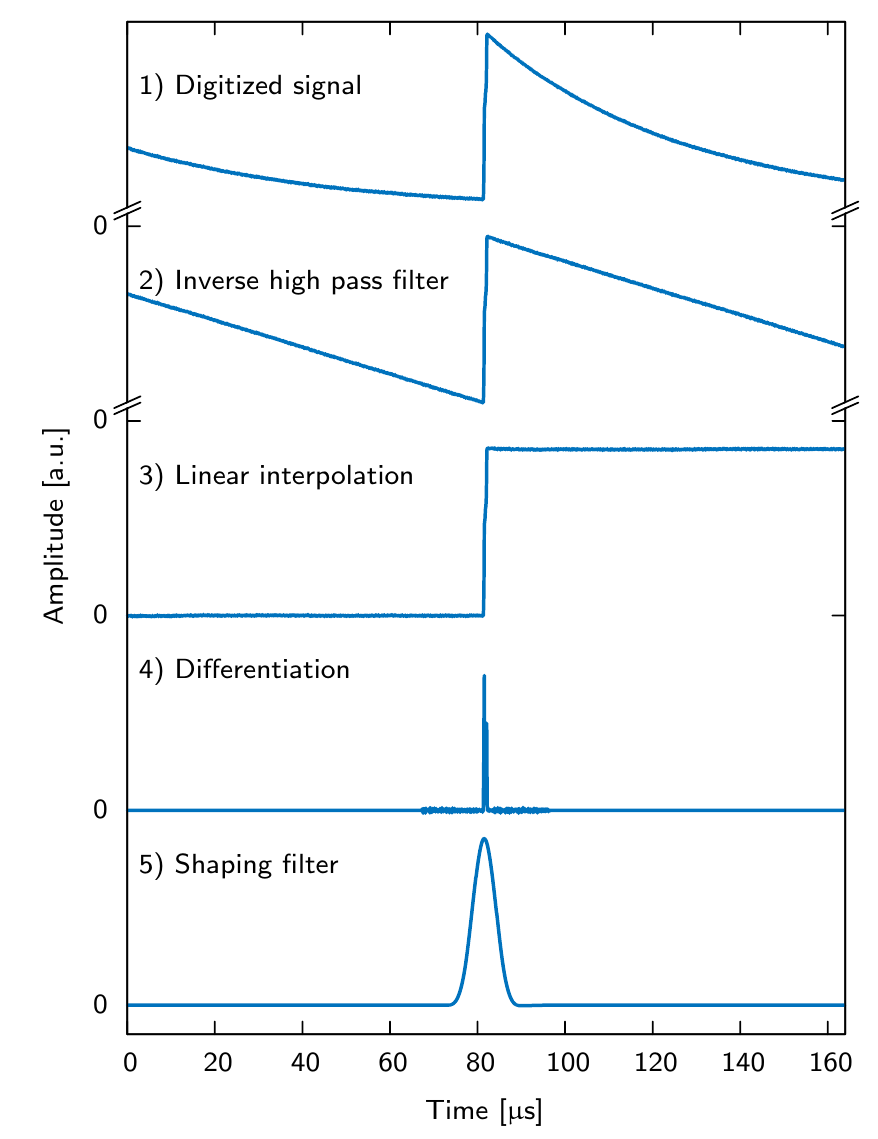}
\caption{\label{fig:filterpulse} The different steps of the pulse height reconstruction applied to a signal with a pileup. The three initial steps illustrate the pole-zero cancellation. It is followed by a differentiation and the shaping filter (pseudo-Gaussian with a shaping time of 6\,$\upmu$s). During the differentiation the analysis range is reduced to a window around the pulse that is wide enough to not considerably affect the shaping filter's performance.}
\end{figure}
The different steps of the pulse height (energy) reconstruction in GEANA are illustrated in Fig.~\ref{fig:filterpulse} and discussed in the following sections. First, the decay introduced by the RC-feedback circuit of the preamplifier in the recorded signal (charge pulse) is compensated with a pole-zero cancellation. This signal is differentiated (current pulse) and can be corrected for ballistic deficits with a nonlinear, irreversible filter, the multi-site event cancellation (MSEC). Finally the shaping filter is applied. The shaping filters are based on a combination of three linear \cite[p.~87]{DSP} filters: the high-pass, low-pass, and moving average filters; their recursive implementations are discussed on p.~277 and p.~322 of Ref.~\cite{DSP}.

\subsection{Pole-zero cancellation}
A high-pass filter has the same impact on a step-like pulse as the RC-feedback circuit does. The recursive form of the high-pass filter can be inverted (to calculate the $n$-th output sample from the input samples $x_i$):
\begin{equation}
  y_n=(g\cdot x_n+\sum_{i=0}^nx_i)\cdot m,\quad g=\frac{d}{1-d},\quad m=2\frac{1-d}{1+d},
\end{equation}
with $d=\exp(\Delta/\tau)$, where $\Delta$ is the sampling width and $\tau$ the decay time of the preamplifier. This inverse high-pass filter can be applied to the signal to compensate the decay introduced by the RC-feedback circuit. However, due to an unknown offset of the baseline, the output signal will have a nonzero slope (see second trace of Fig.~\ref{fig:filterpulse}). This slope can be removed by a simple linear interpolation. The interpolation is performed by dividing the baseline (the recorded samples in the range before the charge pulse) into two segments and calculating their averages. From the two averages the offset and slope of the baseline is extracted and subtracted from the entire pulse. As shown in Fig.~\ref{fig:filterpulse}, this procedure corrects also distortions from the decaying tail of proceeding events (pileups).

\subsection{Ballistic deficit correction}
\label{sec:risetime}
\begin{figure}
\centering \includegraphics{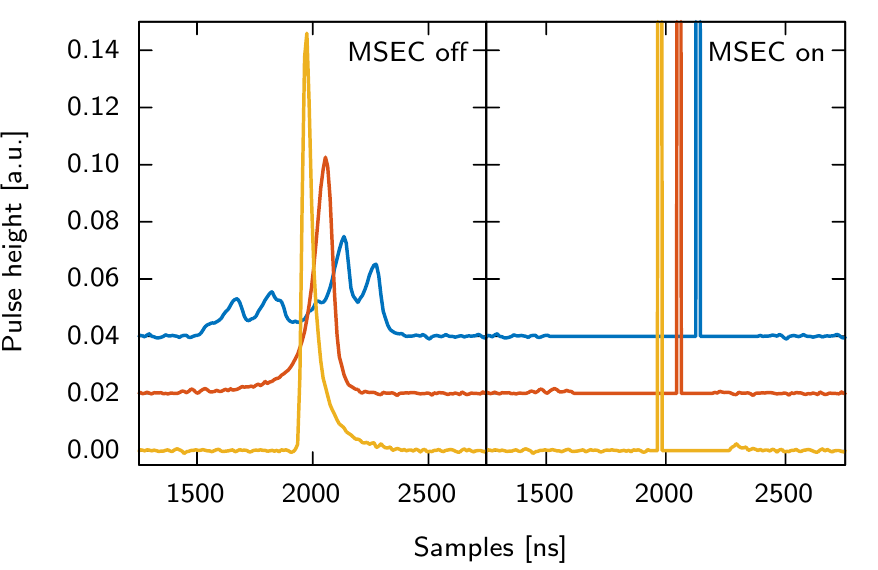}
\caption{\label{fig:msepulse} 
The effect of the multi-site event cancellation (MSEC) method on three different current pulses (normalized). All the contributions around the pulse are added up to a single sample (the size of that samples exceeds the y-range up to roughly 1.)}
\end{figure}
The multi-site event cancellation (MSEC) method has been developed to correct for variability in the pulse shape. It is particularly tailored to correct for events with energy depositions in multiple locations (multi-site event, MSEs) that present particularly strong variations in the charge collection time. It is similar to the method that uses a gated integrator, however, it does not prefilter the signal. Instead, after pole-zero correction and differentiation, the signal is set to zero in a window that stretches over the duration of the charge collection process. The subtracted contributions are then added to the position, where the pulse previously reached its maximum. The width of the window to which the filter is applied is found independently for each individual signal. This is done by searching for the first sample to the left and to the right of the maximal value of the current signal that falls below a certain threshold. The threshold level is extracted from the baseline of the signal. The altered signal is then fed into the shaping filter. The respective procedure is applied to a variety of pulses in Fig.~\ref{fig:msepulse} to show the effect of the MSEC method.

\subsection{Shaping Filters}
\label{sec:filtertype}
The pseudo-Gaussian filter is composed of multiple consecutive moving average filters of the same width \cite[p.~281]{DSP}. The impulse response of the first moving average filter takes the shape of a rectangular function, the second moving average filter changes this into a triangular function. Each further moving average filter transforms the impulse response closer to a Gaussian shape. In this work the pseudo-Gaussian filter is composed of four consecutive moving average filters.

The standard trapezoidal filter is constructed with two moving average filters of different width. The difference in width of the two moving average filters defines the length of the flat-top of the filters output signal. The pulse height ideally is reconstructed at a constant position towards the end of the flat-top \cite{jordanov}. However, such a location is not easily found for a pulse with a variable time profile. This ambiguity is bypassed by applying two additional moving average filters (of half the size of the flat-top) to change the flat-top to a rounded-top and form a clear, distinguished maximum. The width of the rounded-top is defined through the width of the initial flat-top portion (the difference in the width of the two initial moving average filters).

The standard cusp filter can be constructed through a bidirectional \cite[p.~330]{DSP} low-pass filter. In this publication an additional moving average filter flattens the peak of the traditional impulse response. This creates a round-top shape, similar to the one described in Ref.~\cite{Kalinin}. This is better suited to the finite charge collection time of HPGe signals. The width of the rounded-top is defined by the width of the additional moving average filter. The cusp filter optimally extends to infinity, however, in a discrete implementation its width must be limited. Thus, the implemented version is applied in the range of $\pm5$ times the shaping time around the charge pulse.

The altered versions of the two filters will be referred to as rounded-top trapezoidal and rounded-top cusp filter.

\subsection{Evaluation of the energy resolution}
\label{sec:resolution}
\begin{figure}
\centering \includegraphics{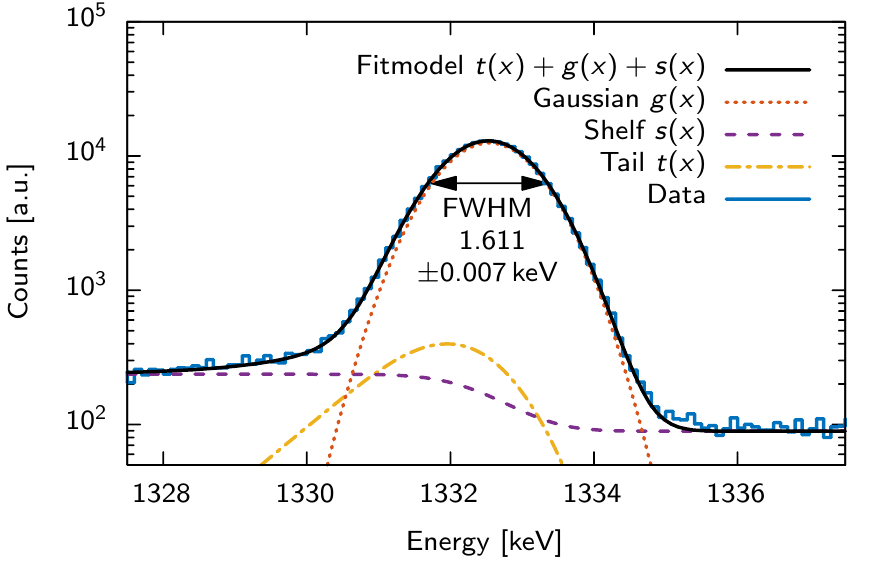}
\caption{\label{fig:fitfunc} The 1332.5\,keV $\gamma$-ray line of $^{60}$Co (pseudo-Gaussian filter, shaping time of 6\,$\upmu$s, ballistic deficits corrected) and the different components of the peak shape model. The uncertainty of the FWHM considers only the uncertainty of the fitting procedure.}
\end{figure}
Statistical fluctuations in the number of produced charge carriers and electronic noise broaden a $\gamma$-ray line's peak into a Gaussian shape. The width of the broadening is measured through the energy resolution, which is a fundamental characteristic of a radiation detector. 

There are different ways to define the energy resolution and it is crucial to clearly define the respective procedure. Within this publication the energy resolution is evaluated by fitting a function (model) to the shape of the peak in the pulse height spectrum by using the Levenberg-Marquardt technique \cite{mpfit}. The model \cite{Campbell1997297} consists of a Gaussian peak $g(x)$, a step-like shelf $s(x)$, and a Hypermet function $t(x)$. The latter is added to account for a possible tail, which decays exponentially below the peak's centroid and is produced by incomplete charge collection and ballistic deficits: 
\begin{align}
	g(x)&=\frac{S_g}{\sqrt{2\pi}\sigma}\exp\left(\frac{-(x-x_0)^2}{2\sigma^2}\right),
 \\
\begin{split}
	t(x)&=\frac{S_t}{2\beta}\exp\left(\frac{x-x_0}{\beta}+\frac{\sigma^2}{2\beta^2}\right) \\ & \qquad \cdot\mathrm{erfc}\left(\frac{x-x_0}{\sqrt{2}\sigma}+\frac{\sigma}{\sqrt{2}\beta}\right),
\end{split}
\\
 	s(x)&=\frac{A}{2}\cdot\mathrm{erfc}\left(\frac{x-x_0}{\sqrt{2}\sigma}\right)+B.
\end{align}
The variables $\sigma$, $x_0$, $\beta$, the number of events in the peak $S_g$ and the tail $S_t$, and the two amplitudes $A$, $B$ are free parameters of the model. After completing the fitting procedure, the energy resolution is calculated from the model as the full-width at half-maximum (FWHM) value of the the Gaussian peak and the decaying tail, i.e.\ $g(x)+t(x)$. Fig.~\ref{fig:fitfunc} shows the $^{60}$Co line at 1332.5\,keV together with the model described in this section.

\section{Setup}
\label{sec:msetup}
To examine the functionality and performance of the methods used in GEANA, data were collected with a standard Canberra Broad Energy Germanium (BEGe) detector \cite{canberra}. The detector is made from p-type germanium, has a mass of 762\,g and a cylindrical form with a radius of 3.6\,cm and a height of 3.5\,cm. A small read-out electrode is embedded in one base of the detector and most of the remaining surface is covered by a high voltage electrode. The BEGe detector is mounted in a 7500SL vertical dipstick vacuum cryostat, which contains a 2002CSL charge sensitive preamplifier ($\sim 50\,\upmu$s decay time). The output of the preamplifier is amplified to fit the dynamic range of the analog to digital converter. The acquisition system consists of a Struck SIS3301 VME flash analog-to-digital converter (FADC) \cite{struck}. The FADC module accommodates up to eight input channels and is equipped with an anti-aliasing bandwidth filter. The signals are sampled in 40\,ns steps and quantized with 16\,bit resolution. The length of the signal is defined to be 164\,$\upmu$s and the rising edge of the pulse is in the center of the sampled time interval.

Two measurements were taken to test the filters, one with a dynamic energy range up to 1750\,keV using a variety of uncollimated radioactive sources ($^{241}$Am, $^{133}$Ba, $^{134}$Cs, $^{137}$Cs, $^{60}$Co, $^{152}$Eu, $^{228}$Th) and one with an energy range up to 3000\,keV with a uncollimated $^{228}$Th source.

\section{Results}
\label{sec:valid}
The $^{60}$Co $\gamma$-ray line at 1332.5\,keV is a standard $\gamma$-ray line in $\gamma$-ray spectroscopy and is used for most of the discussion. Where useful, lines at low energies (59.5\,keV of $^{241}$Am, 121.8\,keV of $^{152}$Eu) and at high energies (2614.5\,keV of $^{228}$Th) are also studied to show differences in a broad energy range. Unless otherwise stated, the standard configuration is the pseudo-Gaussian filter with a shaping time of 6\,$\upmu$s and MSEC enabled. The following discussion is valid for BEGe-like detectors. Nevertheless, the MSEC method can also be used to correct for variations in the pulse shape found in other detector types.

\subsection{Performance of the MSEC method}
\begin{figure}
\centering \includegraphics{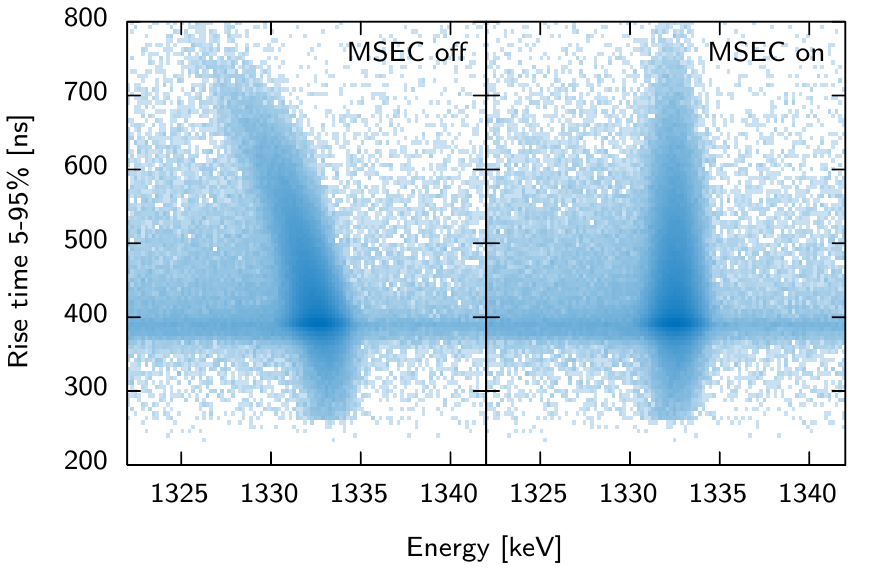}
\caption{\label{fig:risetime} 
A scatter plot of the reconstructed energy and the rise time distribution around the 1332.5\,keV $\gamma$-ray line of $^{60}$Co with the multi-site event cancellation (MSEC) enabled and disabled (pseudo-Gaussian filter, shaping time of 6\,$\upmu$s).}
\end{figure}
\begin{figure}
\centering \includegraphics{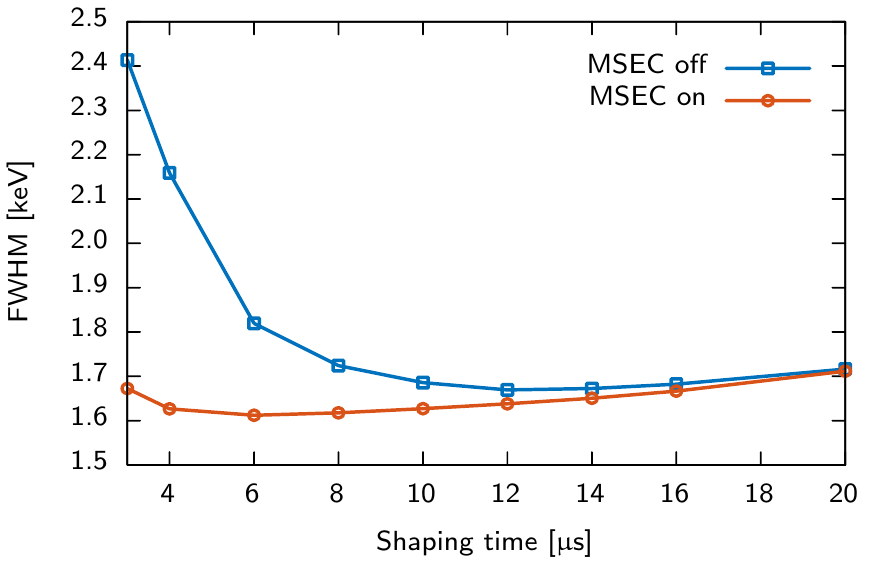}
\caption{\label{fig:shaping} The energy resolution (FWHM) of the 1332.5\,keV $\gamma$-ray line of $^{60}$Co obtained with the pseudo-Gaussian filter for different shaping times. The uncertainties are not drawn; their values are less than 0.5\% and thus not visible.}
\end{figure}
\begin{figure}
\centering \includegraphics{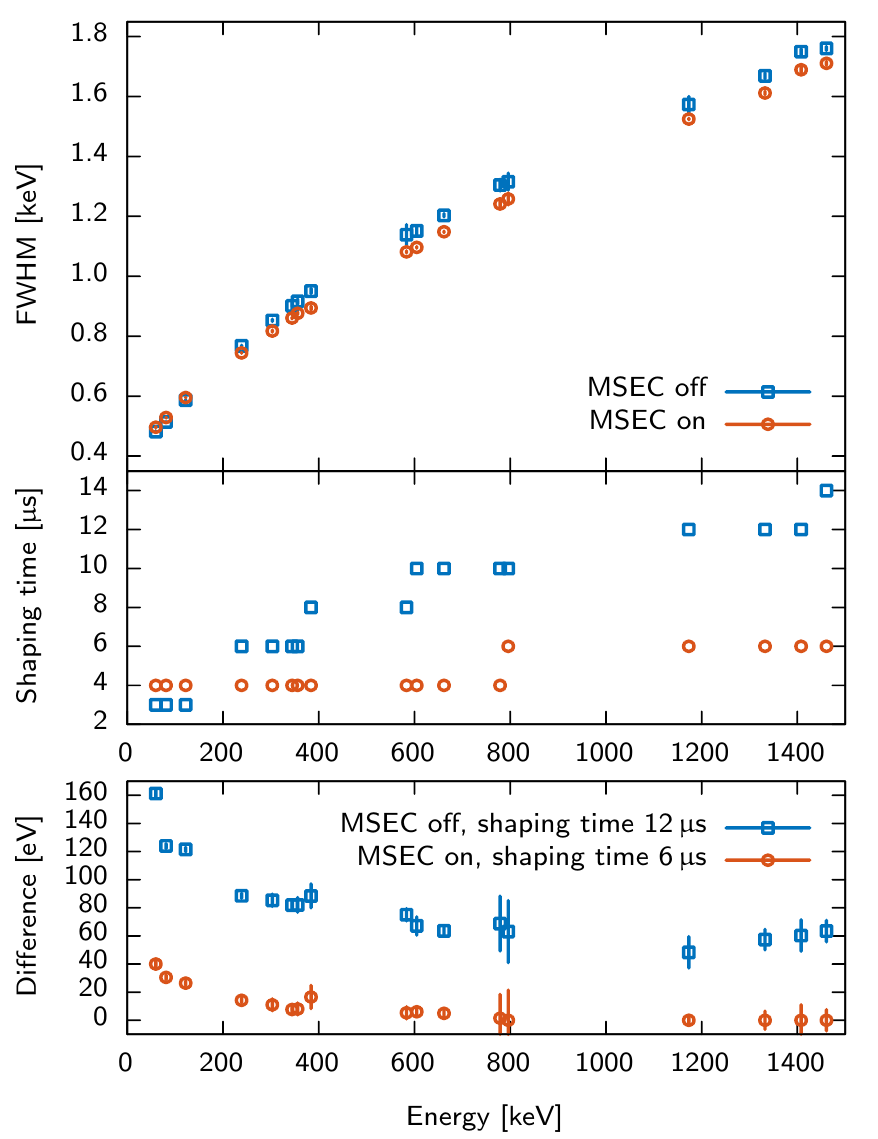}
\caption{\label{fig:dependency} The best energy resolutions (FWHM) found individually for each $\gamma$-ray line with the pseudo-Gaussian filter (top) and the corresponding optimum shaping time (middle). The difference in energy resolution of an analysis with a constant shaping time optimized for the 1332.5\,keV $\gamma$-ray line and the best energy resolution found for an individual $\gamma$-ray line is represented in the bottom.}
\end{figure}
Fig.~\ref{fig:risetime} shows a scatter plot of the energy reconstructed with a pseudo-Gaussian filter of 6$\upmu$s shaping time and the rise time for the region around the 1332.5\,keV $\gamma$-ray line of $^{60}$Co. In the left panel ballistic deficits are not corrected. The shaping filter therefore reconstructs a lower energy for events with a long rise time. In the right panel ballistic deficits are corrected with the MSEC method, resulting in the energy peak being reconstructed for any rise time value in a much narrower energy interval.

In addition, the shaping time of the pseudo-Gaussian filter can be varied. Fig.~\ref{fig:shaping} illustrates, how the energy resolution depends on both, the shaping time and the MSEC method at 1332.5\,keV. The resolution is dominated by voltage noise at short shaping times and by current noise at long shaping times. The optimum shaping time is found at the position where the energy resolution is minimal and the two contributions are equal (the noise corner). A similar behavior can be observed for any $\gamma$-ray line in the pulse height spectrum.

Effects from ballistic deficits further broaden the energy resolution and shift the optimum shaping time and the minimal reachable energy resolution to higher values. The MSEC method corrects this effect and improves the optimum shaping time from 12\,$\upmu$s to 6\,$\upmu$s at 1332.5\,keV (see Fig.~\ref{fig:shaping}). The energy resolution at that energy additionally improves by roughly 4\% from $(1.669\pm0.009)$\,keV to $(1.612\pm0.006)$\,keV. A short shaping time allows for a measurement at higher count rates. Next to improving the energy resolution, the MSEC correction also reduces the optimum shaping time and increases the maximum event rate that can be processed.

\subsection{Energy dependence of the MSEC method}
The top panel of Fig.~\ref{fig:dependency} presents the curve that relates the energy resolution of the most prominent peaks to the energy. At low energies the resolution is primarily dominated by electronic noise. With increased energy the contribution from statistical fluctuations in the number of charge carriers increases and dominates the energy resolution. MSEs from Compton scattering are the most important reason for variations in the pulse shape in BEGe detectors and thus the most important cause of ballistic deficits. In germanium, Compton scattering is surpassed by photoelectric absorption as the dominant interaction channel below $100-200$\,keV. With the disappearance of Compton scattering at low energies the effects caused by ballistic deficits also disappear. 

Above $\sim400$\,keV the MSEC method improves the energy resolution on average by roughly 50\,eV. The improvement is not energy dependent. Below $\sim400$\,keV the improvement quickly decreases and below 200\,keV the MSEC method actually deteriorates the energy resolution. The reason for this is that the summing operation of the MSEC method allows more noise to pass through the shaping filter in respect to the situation where the MSEC method is not applied.

If the ballistic deficits are not corrected then the optimum shaping time is energy dependent. Ballistic deficits grow with increasing energy which in turns increases the optimum shaping time to compensate the ballistic deficits (see Fig.~\ref{fig:shaping}) at these energies.
This effect is illustrated in the central panel of Fig.~\ref{fig:dependency}. The effect is reduced considerably if the MSEC is enabled, but not entirely removed. This suggest that the MSEC method is capable of removing an important fraction of the effects caused by ballistic deficits but not all of them.

In a typical analysis the shaping time optimization is not performed for each peak individually, instead, a single shaping time is chosen for analyzing the entire energy spectrum. Most of the peaks therefore are not reconstructed with the optimum shaping time and the best energy resolution is not always obtained. An analysis with a constant shaping time has been performed for both situations, MSEC disabled and enabled. The shaping time that was used to reconstruct the spectrum was selected by finding the optimum value at the 1332.5\,keV $\gamma$-ray line of $^{60}$Co (12\,$\upmu$s for the MSEC disable and 6\,$\upmu$s for the MSEC enabled). The energy resolutions from a spectrum obtained with only a single shaping time are then compared to the energy resolutions that were found when the shaping time was optimized (by finding the noise corner) at each peak individually. The bottom panel of Fig.~\ref{fig:dependency} presents the difference of the energy resolution obtained with a single shaping time and the value found at the noise corner of each line. The maximal deviation when the MSEC method is disabled is roughly 150\,eV. When the MSEC method is enabled the loss in energy resolution is below 40\,eV. This is also valid if the peaks from $^{228}$Th, which are above 1500\,keV, are included. Accordingly, with the pseudo-Gaussian filter and the MSEC method enabled it is possible to analyze the entire energy range from 59.5\,keV up to 2614.5\,keV with a single shaping time without significant losses in the energy resolution. Therefore, the MSEC method is particularly important for an analysis that only uses a single shaping time over a large range of energies. This is the case for most applications in $\gamma$-ray spectroscopy.

\subsection{Trapezoidal and cusp filter}
\begin{figure}
\centering \includegraphics{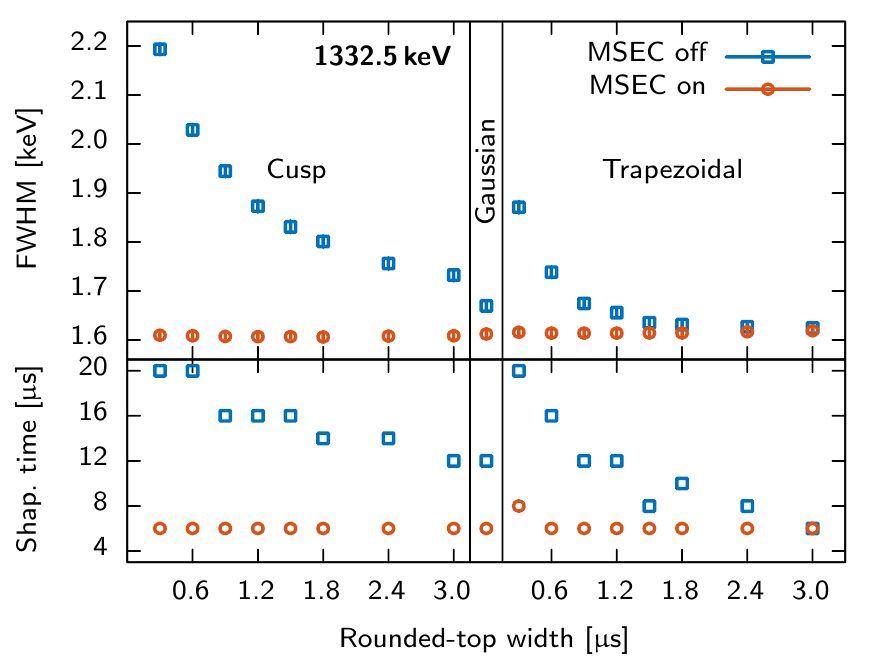}
\centering \includegraphics{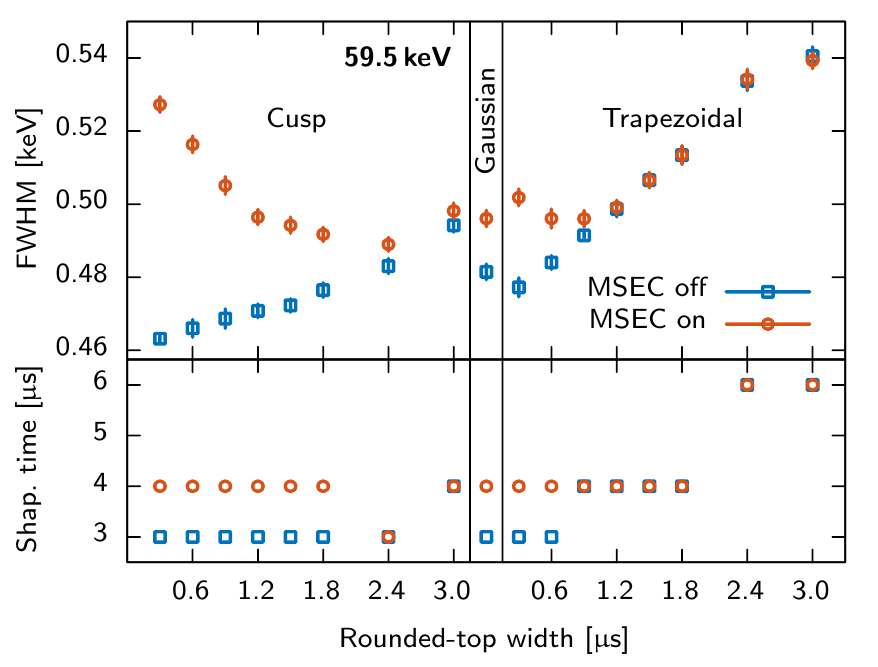}
\caption{\label{fig:filter} The best energy resolution and the optimum shaping time of the 1332.5\,keV (top) and the 59.5\,keV (bottom) $\gamma$-ray lines for the rounded-top cusp and rounded-top trapezoidal filter with various rounded-top widths (see Sec.~\ref{sec:filtertype}). For comparison, the pseudo-Gaussian filter is shown.}
\end{figure}

The MSEC method can also be combined with other filters such as the rounded-top cusp and the rounded-top trapezoidal filter introduced in Sec.~\ref{sec:filtertype}. 

At 1332.5\,keV, with MSEC enabled the two filters have a similar performance to the pseudo-Gaussian filter for all rounded-top widths (see the upper plot of Fig.~\ref{fig:filter}, MSEC on). On average, the tested filters found a resolution of 1.611\,keV with a standard deviation of 0.004\,keV. Thus, the differences between the filters are within the uncertainties of the respective fits. Also the same optimum shaping time is found within a narrow range of values. Where ballistic deficits dominate, the choice of a specific filter does not matter much as long as MSEC is enabled.

The picture is different if ballistic deficits are not corrected and the real sensitivity of the filters to these effects appears (see the upper plot of Fig.~\ref{fig:filter}, MSEC off). At 1332.5\,keV with MSEC disabled filters with a wide rounded-top perform better as more of the ballistic deficits are removed. Moreover, all filters have a tendency towards higher shaping times as this also helps to remove ballistic deficits. The rounded-top cusp filter is least suited to be used in the presence of strong ballistic deficit effects and even with a large rounded-top width it is not capable of matching the performance of the other two filters. The pseudo-Gaussian filter performs slightly better, however, only the rounded-top trapezoidal filter reaches a performance comparable to the situation with MSEC enabled. This requires a large rounded-top width but allows for a much lower shaping time than observed with the other filters.

At 59.5\,keV the situation is quite different (see bottom plot of Fig.~\ref{fig:filter}) as ballistic deficits are not present. Similar to the pseudo-Gaussian filter, the energy resolution found with the two other filters is worse if the MSEC method is enabled. When the MSEC method is disabled, at these energies the rounded-top cusp filter reaches its full noise reduction potential and performs better than the other two filters. Both filters with an adjustable rounded-top perform best when the rounded-top width is small.

The filter's behavior is tightly linked with both, the presence of ballistic deficits and their performance in reducing noise. When ballistic deficits are present it is vital to correct for these effects. Best suited to do this is the MSEC method, but a rounded-top trapezoidal filter can also be used. In the absence of ballistic deficits (i.e.\ at low energies) a cusp filter (with a small rounded-top width) reduces the noise more efficiently than the other two filters and thus performs better. However, it is important to note, that because of its slow falling form, the cusp filter requires a much larger signal range and is not suitable to high count rate measurements.

\subsection{Comparison with other measurements}

Some of the evaluated energy resolutions can be compared to the values provided by the manufacturer of the BEGe detector. The manufacturer quoted an energy resolution of 1.77\,keV at 1332.5\,keV, which is higher than the averaged value of $(1.611\pm0.004)$\,keV found in the previous section. The value measured here is also better than the resolution of $\sim1.63$\,keV found in Ref.~\cite{Harkness2014} and the best value of $(1.65\pm0.01)$\,keV found in Ref.~\cite{Agostini2015}. Both publications used a very similar detector for their studies. The lowest FWHM value measured in this study was $(0.582\pm0.001)$\,keV for the $^{152}$Eu $\gamma$-ray line at 121.8\,keV. This value can be compared to the energy resolution of 0.66\,keV measured by the manufacturer at an energy of 122.1\,keV with a $^{57}$Co source. These comparisons show that the implemented filter methods perform well. 

The best energy resolution obtained for the $^{241}$Am $\gamma$-ray line at 59.5\,keV is $(0.463\pm0.001)$\,keV and the $^{208}$Tl $\gamma$-ray line at 2614.5\,keV is $(2.315\pm0.008)$\,keV. The indicated uncertainties are the standard deviation of the fit parameters.

\section{Conclusion}
Important aspects of digital signal processing techniques used in radiation detectors, such as different shaping filters and a novel method to correct ballistic deficits have been examined for a Canberra Broad Energy Germanium detector (BEGe). With the multi-site event cancellation (MSEC) method enabled all filters lead to comparable results at 1332.5\,keV, the differences are within the uncertainties. The trapezoidal filter with a large round-top has the best performance at 1332.5\,keV if ballistic deficits are not corrected. At energies below $\sim200$\,keV a ballistic deficit correction is not required, instead, it is more beneficial to use a cusp like filter at these energies to find the best signal-to-noise ratio. 

The optimum shaping time for a specific $\gamma$-ray line has been found to be energy dependent. But even an analysis with a single shaping time over a large energy range did deviate by less than 40\,eV (pseudo-Gaussian filter) from the optimal value, provided the MSEC is applied. Energy peaks above a few hundred keV are dominated by Compton scattered events and a correction for ballistic deficits is necessary to avoid an even larger deterioration of energy resolution.

These filters are implemented into GEANA. The software can filter up to 30'000\,events per second (on a single Xeon E5-2660 2.2GHz CPU), if the data can be supplied with an I/O of over 300\,MB/s. Furthermore, the software also performs a fully automated energy calibration and filter optimization. GEANA has already successfully been used as an independent cross check in the Phase I data release \cite{PhysRevLett.111.122503} of GERDA. No discrepancies have been found in the measured energy range, which reaches up to 7\,MeV. Furthermore, it was extensively used for the pulse shape analysis of semi-coaxial detectors in that data release \cite{kirsch}. In the second data collection phase of the experiment, GEANA is being adopted to analyze data gathered from photomultiplier tubes and silicon photomultipliers that are part of the new liquid argon instrumentation \cite{gerdaupgrade}.

\section{Acknowledgment}
The authors are grateful to K.T.\ Kn\"opfle, M.\ Lindner, W.\ Maneschg, and B.\ Schwingenheuer for stimulating discussions and their support and Canberra, Olen for its prolific cooperation during the construction of the detector used in this work.

\section{References}
\bibliographystyle{./elsarticle-num-names}
\bibliography{references}

\end{document}